\documentclass[acmsmall,screen,nonacm]{acmart}

\usepackage{booktabs}
\usepackage{multirow}
\usepackage{graphicx}
\usepackage{xcolor}
\usepackage{tcolorbox}
\usepackage{pifont}
\newtcolorbox{titleEnv}{
colframe=black!80,
colback=gray!10,
fonttitle=\bfseries,
coltitle=black,
left=3pt,
right=3pt,
top=3pt,
bottom=3pt,
boxrule=0.4mm,
arc=3mm
}
\usepackage{graphicx}
\usepackage{booktabs} 
\usepackage{multicol}
\usepackage{multirow}
\usepackage{appendix}
\usepackage{setspace}
\usepackage{makecell}
\usepackage{url}

\title{Characterizing Security Effects of OSS Vulnerabilities in Agent Systems}

\author{Yu Ji}
\affiliation{
  \institution{Huazhong University of Science and Technology}
  \city{Wuhan}
  \country{China}
}
\email{jiyu@hust.edu.cn}

\author{Yang Wei}
\affiliation{
  \institution{Huazhong University of Science and Technology}
  \city{Wuhan}
  \country{China}
}
\email{weiyang\_@hust.edu.cn}

\author{Yutao Hu}
\affiliation{
  \institution{Huazhong University of Science and Technology}
  \city{Wuhan}
  \country{China}
}
\email{yutaohu@hust.edu.cn}

\author{Haojun Zhao}
\affiliation{
  \institution{Huazhong University of Science and Technology}
  \city{Wuhan}
  \country{China}
}
\email{zhaohaojun@hust.edu.cn}

\author{Yueming Wu}
\affiliation{
  \institution{Huazhong University of Science and Technology}
  \city{Wuhan}
  \country{China}
}
\email{wuyueming21@gmail.com}

\author{Deqing Zou}
\affiliation{
  \institution{Cyberspace Security University of China}
  \city{Wuhan}
  \country{China}
}
\email{deqingzou@hust.edu.cn}

\setcopyright{none}

\begin{document}

\begin{abstract}
Software agents increasingly depend on open source components when executing tools and interacting with external systems. Security flaws in these dependencies may therefore influence more than the software process in which they occur: their consequences can be carried through tool outputs, agent state, and information subsequently exposed to the model. Determining whether such a consequence is actually realized in a particular execution, and where its influence stops within the agent system, remains challenging.
We investigate how known OSS vulnerabilities behave when exercised as part of agent workflows. Our study reveals recurring patterns in the way security-relevant consequences emerge and propagate across runtime layers. Building on these observations, we use vulnerability-aware semantic information, differential executions of vulnerable and corrected software, and runtime provenance spanning multiple layers to determine whether a vulnerability produces an observable security effect and to identify the furthest layer at which that effect remains manifested.
Our evaluation shows that this approach can accurately distinguish realized vulnerability effects and determine their manifestation boundaries across a diverse collection of vulnerability scenarios. We additionally apply the analysis to documented workflows in a real-world agent framework and uncover multiple security effects originating from known vulnerabilities in its OSS dependencies. These results highlight the importance of reasoning about vulnerable dependencies in terms of their execution-level consequences rather than vulnerability presence alone.
\end{abstract}

\ccsdesc[500]{Security and privacy~Software security engineering}
\ccsdesc[500]{Software and its engineering~Dynamic analysis}

\keywords{LLM Agents, OSS Vulnerabilities,
Realized Vulnerability Effects, Cross-Layer Runtime Provenance,
Manifestation Boundaries}

\maketitle

\section{Introduction}
\label{sec:introduction}

\noindent \textbf{How Do OSS Vulnerabilities Manifest in Agents?}
As agents become increasingly capable, completing complex tasks increasingly relies on \emph{open source software} (OSS) to interact with external environments, including network services, files, databases, and executable code. Agent Hosts, Tools, MCP Servers, and their underlying protocol libraries, parsers, HTTP clients, and data-storage components together form the software stack that executes agent actions \cite{hou2026mcplandscape}.
These components may themselves contain publicly disclosed software vulnerabilities \cite{hasan2026mcpfirstglance}.

In agent systems, the effects of OSS vulnerabilities may extend beyond the conventional software runtime.
They can pass through Tool returns and Host processing, reach model-visible Observations, and become inputs to subsequent agent decisions
\cite{zhan2024injecagent,debenedetti2024agentdojo,jia2025taskshield}.
For example, a vulnerability may cause an unauthorized file read, unintended credential use, or improper resource modification at runtime.
The resulting data or state may then be returned by a Tool, transformed by the Host, and exposed to the model.
Across this process, the concrete operation, state, or value carrying the vulnerability effect may change, while the security-relevant consequence may remain represented.

This raises a fundamental question for analyzing known OSS vulnerabilities in agents. What security-relevant effect does a target vulnerability actually cause in a concrete agent execution, and how far through the agent stack does that effect remain manifest as its representation changes?

\noindent \textbf{What Is Missing from Existing Analysis?}
The core difficulty is connecting a vulnerability-induced runtime change to its downstream manifestation in the agent execution. This connection cannot always be established by tracking the original runtime value. A vulnerability may alter a resource state, authorization behavior, or external response. The resulting Tool output or Observation may no longer contain the original value, yet may still reflect the same security consequence.

Existing analyses address different parts of this problem.
OSS vulnerability analysis can identify vulnerable dependencies \cite{ponta2020detection,duan2017osspolice}, determine whether vulnerability-related code is reachable \cite{ponta2018beyond,hu2026vulsca}, and determine whether vulnerable constructs are actually exercised at runtime \cite{ponta2020detection,ponta2018beyond}.
These techniques establish whether a vulnerability is present, reachable, or exercised in the underlying software.
Data-flow and taint analysis can further track information dependencies
\cite{he2026taintp2x,hou2026mcpbiflow}.
However, once a vulnerability changes a resource state, authorization decision,
external response, or other runtime outcome, later Agent representations may no
longer preserve a direct correspondence to the original operation, state, or value.
Existing OSS-centric analyses therefore do not directly determine whether the same vulnerability-attributable effect remains manifest after such transformations.

Agent security analysis starts from the other end of the execution.
Existing work examines security-relevant content in Tool outputs and model-visible Observations \cite{hou2026mcpbiflow,radosevich2025mcpsafetyaudit,greshake2023indirect,wang2025agentvigil}, as well as subsequent agent behavior \cite{ruan2024toolemu,zhu2025melon,an2025ipiguard}.
However, it typically does not trace these Agent-level phenomena back to the
realized runtime effect of a specific known OSS vulnerability.
This leaves a missing link between vulnerability-specific runtime effects and their downstream manifestations in the Agent execution.

\noindent \textbf{What Must Be Identified?}
We therefore decompose the problem into two analyses. {Vulnerability attribution} identifies the security-relevant runtime change caused by the target vulnerability in a given execution. {Effect manifestation} determines the furthest execution boundary at which that attributed effect remains manifest.

We accordingly ask the following research question.
\emph{Given an agent execution, what security-relevant effect is actually caused by a target OSS vulnerability, and what is the furthest execution boundary at which that effect remains manifest?}

\noindent \textbf{Preliminary Study.}
We first conduct a preliminary study of 52 vulnerability cases. Across these cases, we observe a recurring effect structure. A {vulnerability-attributable root runtime change} acts on specific objects or states and may further produce security-relevant consequences. We also observe different manifestation depths. Some effects remain within the underlying runtime, while others enter the Tool-return path or reach the model-visible Observation.
Based on these observations, we define a {realized vulnerability effect} as the security-relevant runtime change caused by the target vulnerability, together with its affected objects or states and any resulting security-relevant consequences. We define the furthest execution boundary at which the effect remains manifest as its {maximum manifestation boundary}.

\noindent \textbf{\textsc{Oscar}.}
Based on this abstraction, we present \textsc{Oscar} (\textbf{O}pen-\textbf{S}ource \textbf{C}VE \textbf{A}nalysis for Agent \textbf{R}untime), a framework for identifying realized vulnerability effects and localizing their maximum manifestation boundaries.
Given a target vulnerability, an agent task, and its input, \textsc{Oscar} extracts {trigger semantics} and {patch semantics} from public vulnerability information and the corresponding patch. It then executes the vulnerable and fixed versions under equivalent conditions. \textsc{Oscar} combines paired execution differences with vulnerability-specific semantics to identify the root runtime change attributable to the target vulnerability. It then uses runtime provenance across the {Program, Resource, MCP, Host, and Agent} layers to recover the affected objects or states, resulting security-relevant consequences, and the maximum manifestation boundary.

\noindent \textbf{Evaluation.}
We evaluate \textsc{Oscar} on an independently constructed vulnerability test set and real-world agent frameworks. \textsc{Oscar} achieves {96.4\% F1} and {96.0\% accuracy} in realized vulnerability effect identification and correctly localizes 24 of 28 maximum manifestation boundaries.
We further deploy five Agent Hosts in the real-world open-source autonomous research framework {ARIS}. Across these heterogeneous agent runtimes, \textsc{Oscar} identifies vulnerability effects that either remain within the underlying runtime or reach model-visible Observations. Ablation experiments show that both {vulnerability-specific semantics} and {vulnerable/fixed paired execution} are important for accurate vulnerability attribution and effect reconstruction.

\noindent \textbf{Contributions.}
In summary, this paper makes the following contributions.

\noindent$\bullet$ \textbf{Problem characterization.}
We formulate the {cross-layer effect manifestation problem} for known OSS vulnerabilities in agent runtimes. The problem is to determine how an effect caused by a specific vulnerability in the underlying software execution manifests across subsequent agent execution when its original code- or value-level correspondence may no longer be preserved.

\noindent$\bullet$ \textbf{Effect abstraction and analysis.}
Through a preliminary study of 52 vulnerability cases, we characterize the recurring structure of realized vulnerability effects and introduce the {maximum manifestation boundary}. Based on these observations, we design \textsc{Oscar}, which combines vulnerability-specific semantics, paired execution, and cross-layer runtime provenance to identify realized effects and localize their manifestation boundaries.

\noindent$\bullet$ \textbf{Evaluation.}
We evaluate \textsc{Oscar} on an independent vulnerability test set and a real-world multi-Host agent environment. The results show that \textsc{Oscar} accurately identifies OSS vulnerability effects realized in concrete executions and localizes their furthest manifestation boundaries.

\section{Preliminary Study}
\label{sec:preliminary}

Having motivated the need to connect vulnerability-attributable runtime effects with their downstream manifestations, we first conduct a preliminary study to understand how such effects appear in concrete Agent executions. Specifically, we examine the runtime structure of realized vulnerability effects and how far these effects remain manifest as execution proceeds from the underlying software runtime to model-visible Observations. 

\subsection{Problem Definition and Scope}
\label{sec:problem-scope}

We study known OSS vulnerabilities in concrete Agent executions initiated by user tasks and inputs. We assume that the target vulnerability and its affected component have already been identified. Our goal is to determine whether the vulnerability produces a security-relevant effect in the given execution and, if so, how far that effect manifests through the Agent execution stack.

Our analysis is execution-specific. We do not identify vulnerabilities in the Agent stack, generate inputs to trigger them, estimate their triggering likelihood under natural workloads, or consider unknown vulnerabilities. Instead, we focus on the runtime consequences of a known vulnerability under a given Agent execution.
Accordingly, we investigate two research questions:

\noindent $\bullet$ \textbf{RQ1:} What security-relevant effects do known OSS vulnerabilities produce in concrete Agent executions?

\noindent $\bullet$ \textbf{RQ2:} How far do these effects manifest through the Agent execution stack?

\subsection{Study Design}
\label{sec:prelim-method}

We map MCP servers, Agent tools, and their dependencies to known vulnerabilities reported in \emph{Open Source Vulnerabilities} (OSV)~\cite{osv}, yielding 52 vulnerability cases involving 49 OSS projects.
For each case, we construct an input and execution environment that satisfy vulnerability-relevant conditions and execute the complete Agent task.

For each execution, we collect the runtime trace from the initial user input to task completion.
The trace covers Tool invocations, program and resource events, runtime values, Tool Results, Host processing, model-visible Observations, and subsequent Agent Actions.

\noindent\textbf{Runtime Effect Analysis.}
We inspect each trace to identify security-relevant events associated with the target vulnerability and record their concrete operations, affected objects or resources, runtime values, and state transitions.
We then group events with the same runtime semantics across cases and analyze the recurring structure and security characteristics of the resulting effects.

\noindent\textbf{Manifestation Analysis.}
For each realized effect, we trace its representation through the Tool Result, Host processing, and model-visible Observation to determine its maximum manifestation boundary and the mechanisms that limit further manifestation.

Three researchers independently analyze the traces, resolve disagreements to establish a common guideline, and re-examine all cases under this guideline.

\subsection{Structure and Characteristics of Vulnerability Effects}
\label{sec:effect-structure}

Among the 52 study cases, 51 exhibit realized vulnerability effects supported by runtime evidence.
We analyze these cases to determine whether their effects share a common runtime structure.

\begin{figure}[t]
    \centering
    \includegraphics[width=\linewidth]{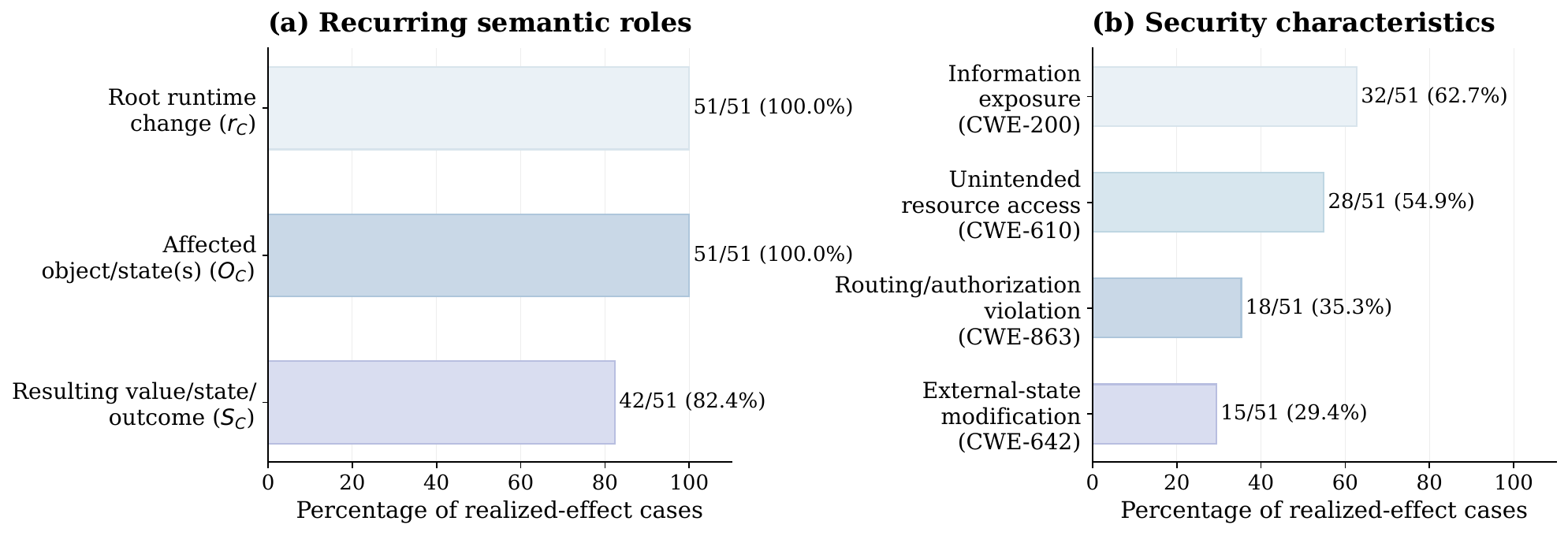}
\caption{Observed runtime structure and security characteristics of realized OSS vulnerability effects.}
    \label{fig:effect-structure}
\end{figure}

As shown in Figure~\ref{fig:effect-structure}(a), all 51 realized effects contain a \emph{root runtime change} and one or more affected \emph{objects or states}.
Among them, 42 cases (82.4\%) further produce a security-relevant value, state, or external outcome.
These recurring semantic roles suggest a common runtime structure across vulnerabilities.
We represent the realized vulnerability effect of case $C$ as
$
E_C=(r_C,O_C,S_C)
$

where $r_C$ is the \emph{root runtime change}, defined as the earliest security-relevant runtime behavior or state transition caused by the target vulnerability.
$O_C$ denotes the objects or states affected by $r_C$, and $S_C$ denotes the security-relevant values, states, or external outcomes subsequently produced.
If $r_C$ acting on $O_C$ already constitutes the realized effect and no further security-relevant outcome is observed, then $S_C=\varnothing$.
For example, in a path-traversal case where the program reads a file outside the permitted directory and obtains its contents, $r_C=\text{out-of-scope file read}$, $O_C={\text{target file}}$, and $S_C={\text{obtained file content}}$.

Figure~\ref{fig:effect-structure}(b) further shows the security characteristics of these effects using CWE terminology.
\emph{Information exposure} (CWE-200), \emph{unintended resource access} (CWE-610), \emph{routing/authorization violation} (CWE-863), and \emph{external-state modification} (CWE-642) occur in 32 (62.7\%), 28 (54.9\%), 18 (35.3\%), and 15 (29.4\%) cases, respectively.
These characteristics are non-exclusive, and a single realized effect may exhibit multiple security consequences.

\begin{titleEnv}
\textit{\textbf{Finding 1}: Realized OSS vulnerability effects exhibit a recurring runtime structure despite diverse security consequences.}
\end{titleEnv}

\subsection{Cross-Layer Manifestation Boundaries}
\label{sec:manifestation-boundary}

We next examine how far each realized effect manifests through the Agent execution stack.
Following the typical execution structure of tool-augmented agents
\cite{yao2023react,schick2023toolformer,ruan2024toolemu},
we inspect the underlying runtime execution, Tool Result, Host processing, and model-visible Observation.
We refer to the execution initiated by one Tool invocation and the processing that produces its corresponding Observation as an \emph{execution round}.

\begin{figure}[t]
    \centering
    \includegraphics[width=\linewidth]{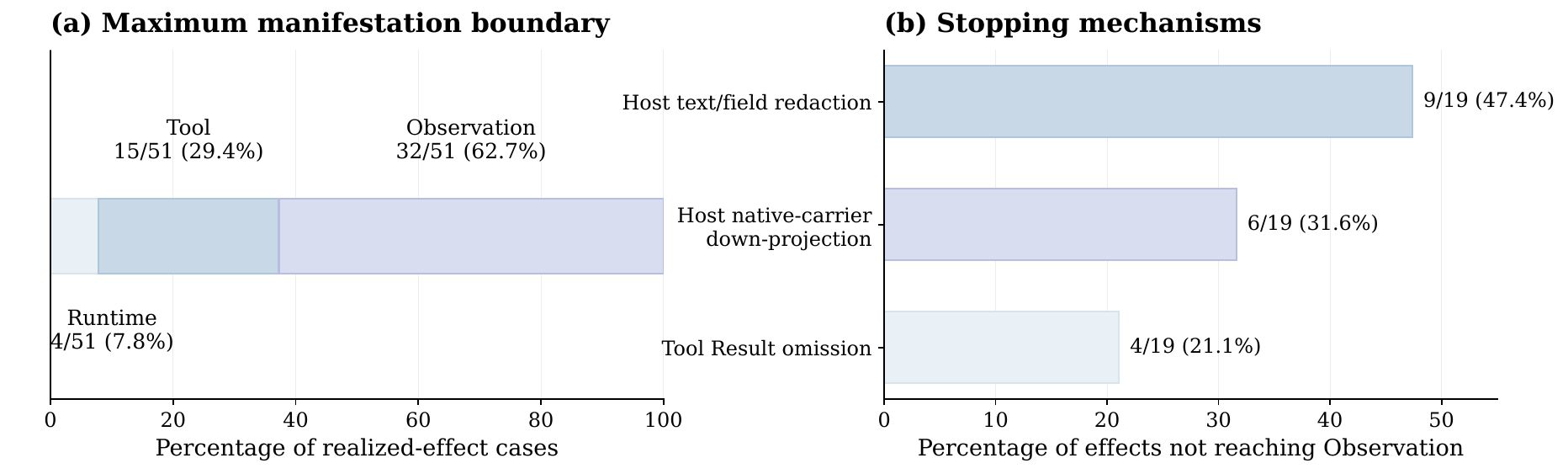}
    \caption{Cross-layer manifestation of realized vulnerability effects.}
    \label{fig:manifestation}
\end{figure}

Across the 51 realized effects, we observe three recurring maximum manifestation boundaries: Runtime, Tool, and Observation.
As shown in Figure~\ref{fig:manifestation}(a), 32 effects (62.7\%) reach the Observation boundary, where the effect remains identifiable in the model-visible Observation.
Another 15 effects (29.4\%) reach the Tool boundary, where a representation enters the Tool-return path but does not reach the Observation after Host processing.
The remaining four effects (7.8\%) remain at the Runtime boundary, where the effect is realized in program execution, external resources, or environmental state, but no representation enters the Tool Result.
These boundaries summarize the furthest scope reached by each effect, with Host processing treated as an intermediate transformation rather than a separate boundary.

For the 19 effects with a Runtime or Tool maximum manifestation boundary, Figure~\ref{fig:manifestation}(b) summarizes the mechanisms that prevent further manifestation.
\emph{Host text/field redaction} occurs in nine cases (47.4\%), where Host processing removes or replaces security-relevant content associated with the effect.
\emph{Host native-carrier down-projection} occurs in six cases (31.6\%), where an image, Blob, or embedded resource no longer preserves the effect after Host-side representation conversion.
\emph{Tool Result omission} occurs in four cases (21.1\%), where an effect realized in the underlying runtime is not represented in the Tool Result.

\begin{titleEnv}
\textit{\textbf{Finding 2}: Realized vulnerability effects exhibit different maximum manifestation boundaries and may terminate before reaching the model-visible Observation.}
\end{titleEnv}

\section{\textsc{Oscar}}
\label{sec:Oscar}

Motivated by the recurring effect structure and manifestation boundaries identified in Section~\ref{sec:preliminary}, we propose \textsc{Oscar} (\textbf{O}pen-\textbf{S}ource \textbf{C}VE \textbf{A}nalysis for Agent \textbf{R}untime) to identify the realized effect of a known OSS vulnerability under a given Agent task and input and localize its maximum manifestation boundary.

\begin{figure}[t]
    \centering
    \includegraphics[width=\linewidth]{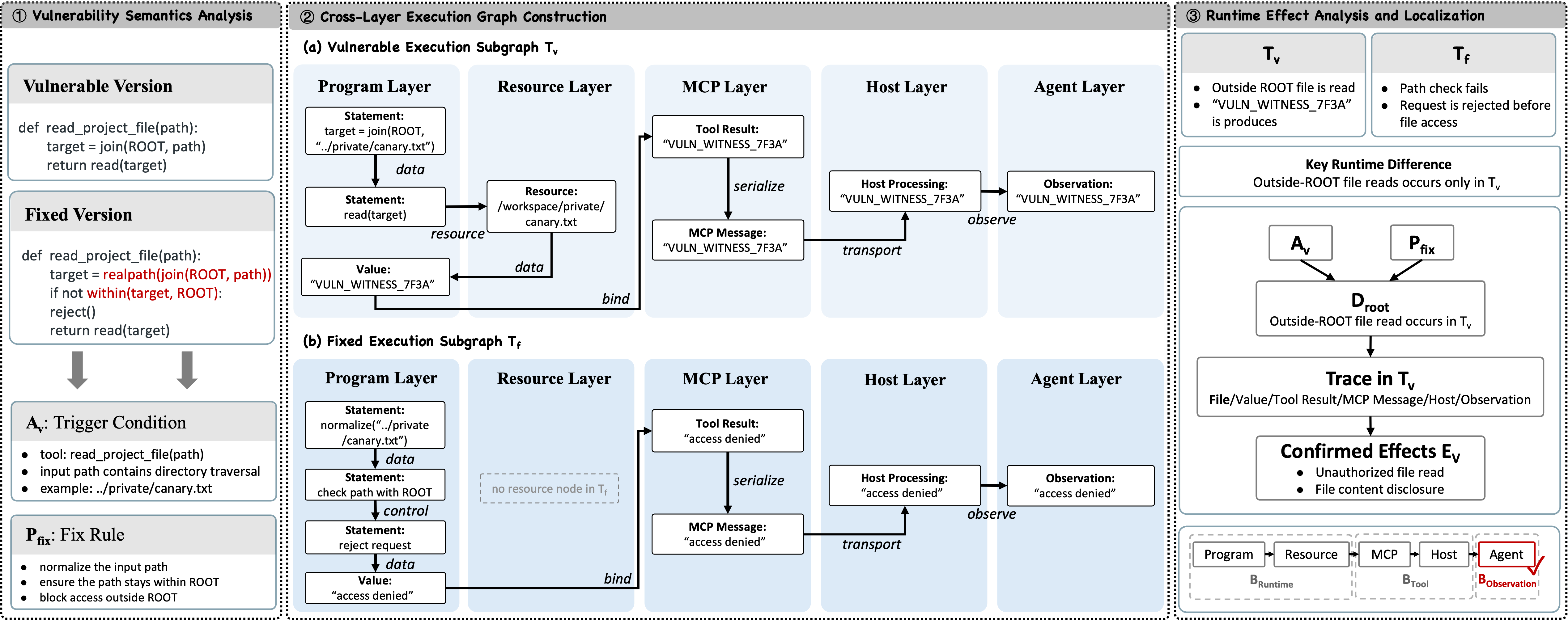}
    \caption{Overview of \textsc{Oscar} with a running example.}
    \label{fig:overview}
\end{figure}

\subsection{System Overview}
\label{sec:Oscar-overview}

Figure~\ref{fig:overview} presents the three-stage workflow of \textsc{Oscar}.
\ding{182} \textbf{Vulnerability Semantic Analysis.}
\textsc{Oscar} extracts vulnerability trigger semantics and patch semantics from public vulnerability information and the corresponding patch to constrain subsequent runtime analysis.
\ding{183} \textbf{Cross-Layer Execution Graph Construction.}
\textsc{Oscar} executes the vulnerable and fixed versions under the same conditions and constructs paired execution graphs spanning the Program, Resource, MCP, Host, and Agent layers to preserve cross-layer runtime provenance.
\ding{184} \textbf{Runtime Effect Analysis and Localization.}
\textsc{Oscar} compares the paired graphs to identify the vulnerability-induced root runtime change, reconstruct the realized effect, and determine its maximum manifestation boundary.

Figure~\ref{fig:overview} also illustrates a running example involving a path-traversal vulnerability in an MCP file-reading Tool. The vulnerable version accesses a file outside the intended directory, whereas the fixed version rejects the same request before file access.

\subsection{Vulnerability Semantic Analysis}
\label{sec:semantic-analysis}

To attribute runtime behavior to the target vulnerability, \textsc{Oscar} extracts vulnerability trigger semantics $A_V$ and patch semantics $P_{\mathit{fix}}$.

\subsubsection{Vulnerability Trigger Semantics.}

$A_V$ captures the input characteristics and runtime conditions required to reach a vulnerability-relevant execution path.
\textsc{Oscar} derives $A_V$ from CVE descriptions, security advisories, PoCs, and related documentation, including the relevant interface, parameters, and trigger conditions.
It is used to construct vulnerability-relevant executions and to constrain subsequent attribution of runtime differences.
Satisfying $A_V$ indicates that the vulnerability-relevant path is reached, but does not by itself establish a realized effect.
In the running example, $A_V$ captures the \texttt{path} parameter and its directory-traversal characteristic, such as \texttt{../private/canary.txt}.

\subsubsection{Patch Semantics.}

$P_{\mathit{fix}}$ captures the vulnerable behavior eliminated or constrained by the fix.
\textsc{Oscar} derives $P_{\mathit{fix}}$ using a vulnerability-fix semantic extraction technique~\cite{gao2025whyhow}, together with vulnerability descriptions, security advisories, and patch-related text.
It is used to determine whether an observed runtime difference corresponds to the behavior addressed by the patch.
In the running example, $P_{\mathit{fix}}$ specifies that access to files outside \texttt{ROOT} must be blocked before file access occurs.

\subsection{Cross-Layer Execution Graph Construction}
\label{sec:graph-construction}

A realized vulnerability effect may originate in program execution, affect external resources, and further propagate through Tool returns and Host processing into the Agent context.
To capture this runtime provenance, \textsc{Oscar} constructs a cross-layer execution graph spanning the Program, Resource, MCP, Host, and Agent layers.

\subsubsection{Execution Graph Definition.}

\textsc{Oscar} models a task-level execution as $G=(N,R)$, where $N$ denotes runtime nodes and $R$ denotes runtime relations.
Table~\ref{tab:graph-nodes} defines the node types and their corresponding execution layers.

\begin{table}[!ht]
\centering
\caption{Node types in the cross-layer execution graph.}
\label{tab:graph-nodes}
\resizebox{0.8\linewidth}{!}{
\begin{tabular}{lll}
\toprule
\textbf{Node Type} & \textbf{Layer} & \textbf{Semantics} \\
\midrule

Function &
\multirow{3}{*}{Program} &
A handler or internal function executed by the OSS component. \\

Statement &
&
A concrete program operation occurring during execution. \\

Value &
&
A runtime value produced, consumed, or propagated during execution. \\
\midrule

Resource &
Resource &
An external resource or state accessed, modified, or created during execution. \\
\midrule

Tool Invocation &
\multirow{4}{*}{MCP} &
A concrete Tool invocation occurring during execution. \\

Tool Parameter &
&
A concrete parameter carried by the corresponding Tool invocation. \\

Tool Result &
&
A concrete result produced by the corresponding Tool execution. \\

MCP Message &
&
A protocol-level message carrying the corresponding Tool Result. \\
\midrule

Host Processing &
Host &
A Host-side transformation applied to the returned Tool result. \\
\midrule

Agent Observation &
\multirow{2}{*}{Agent} &
Model-visible content produced after Host-side processing. \\

Agent Action &
&
An observable Agent action produced from the current Observation. \\

\bottomrule
\end{tabular}
}
\end{table}

Relations in $R$ connect these nodes through invocation, control, data provenance, resource interaction, and cross-layer propagation, as summarized in Table~\ref{tab:graph-relations}.
Among them, Call edges represent invocation only, while returned values are captured by Data or Bind edges.
Resource edges connect program operations to the external resources or states they access or modify.

\begin{table}[!ht]
\centering
\caption{Relation types in the cross-layer execution graph.}
\label{tab:graph-relations}
\resizebox{\linewidth}{!}{
\begin{tabular}{llll}
\toprule
\textbf{Relation Type} & \textbf{Source Node} & \textbf{Target Node} & \textbf{Semantics} \\
\midrule

Call &
Tool Invocation, Function &
Function &
Invocation of a Tool handler or internal function. \\

Control &
Statement &
Statement, Function &
Control of subsequent execution by a condition or check. \\

Data &
Resource, Statement, Function, Value &
Value, Statement, Function, Resource &
Transfer of runtime data between execution entities. \\

Resource &
Statement, Resource &
Resource, Statement &
Interaction between program execution and an external resource. \\

Bind &
Tool Invocation, Tool Parameter, Function, Value &
Tool Parameter, Function, Value, Tool Result &
Binding between Tool-level and program-level runtime entities. \\

Serialize &
Tool Result &
MCP Message &
Serialization of a Tool Result into an MCP message. \\

Transport &
MCP Message &
Host Processing &
Transfer of an MCP message to Host processing. \\

Observe &
Host Processing &
Agent Observation &
Formation of an Agent Observation from Host processing. \\

Action &
Agent Observation, Agent Action &
Agent Action, Tool Invocation &
Transition from Agent Observation to action or Tool invocation. \\

\bottomrule
\end{tabular}
}
\end{table}

Each node denotes a concrete runtime instance rather than a static program entity.
The same function, statement, or resource may therefore appear multiple times when it participates in different execution rounds or repeated operations. Relations are instantiated only for dependencies observed in the current execution, so the graph preserves the concrete provenance needed to connect a vulnerability-related runtime change to the Tool, Host, and Agent entities that subsequently carry its effect.
An Agent task may span multiple execution rounds.
Each round begins with a Tool Invocation and proceeds through program execution, resource interaction, Tool return, Host processing, and Agent Observation.
If the resulting Agent Action triggers another Tool invocation, \textsc{Oscar} continues tracing until task completion.

\subsubsection{Paired Execution Graph Generation.}

\textsc{Oscar} uses the fixed version as an execution-level reference to distinguish vulnerability-induced behavior from normal task behavior.
It executes the vulnerable and fixed versions under the same Tool input, Host configuration, and external runtime environment, producing paired execution graphs $G_v$ and $G_f$.

For each execution, \textsc{Oscar} records the concrete Tool parameters, executed program operations, resource/state interactions, runtime values, Tool returns, Host processing, Agent Observations, and subsequent Agent Actions.
Only events that actually occur are instantiated in the graph.
Thus, the paired executions differ only in the target OSS version at setup, while
$G_v$ and $G_f$ preserve the runtime divergences induced by that version change.

In the running example, the vulnerable version receives \texttt{../private/canary.txt}, accesses \texttt{/workspace/private/canary.txt}, and obtains \texttt{VULN\_WITNESS\_7F3A}.
Thus, $G_v$ contains the out-of-scope file access and its subsequent value propagation.
The fixed version rejects the same input before file access, so these events are absent from $G_f$.

\subsection{Runtime Effect Analysis and Localization}
\label{sec:effect-analysis}

With the paired execution graphs constructed, \textsc{Oscar} next identifies the realized vulnerability effect and localizes its manifestation.
The analysis proceeds from the root runtime change, to effect reconstruction, and finally to maximum manifestation boundary localization.

\subsubsection{Root Runtime Change Identification.}

Paired executions may exhibit many runtime differences, including both vulnerability-induced behavior and downstream changes caused by execution divergence. To identify where the realized effect first emerges, \textsc{Oscar} aligns runtime entities in $G_v$ and $G_f$ with corresponding execution semantics, and compares the aligned graphs to identify differences in runtime events, relations, states, and values. Because graph differences alone do not establish vulnerability causality, \textsc{Oscar} further attributes each candidate difference using $A_V$ and $P_{\mathit{fix}}$: $A_V$ confirms that the execution satisfies the trigger conditions, while $P_{\mathit{fix}}$ determines whether the difference corresponds to behavior eliminated or constrained by the patch.

We distinguish three forms of candidate differences during graph comparison.
An event or relation may occur only in one execution; aligned runtime entities
may carry different values or states; or the same operation may interact with
a different resource or produce a different outcome.
These candidates are identified before vulnerability attribution.
$A_V$ and $P_{\mathrm{fix}}$ are then used to retain only differences that are consistent with both the trigger conditions and the behavior addressed by the patch.
Among the attributed differences, \textsc{Oscar} defines the earliest security-relevant runtime change as the root runtime change $r_V$, which serves as the provenance root of the realized effect. If no such change exists, \textsc{Oscar} reports $E_V=\varnothing$ and stops the analysis. In the running example, \texttt{../private/canary.txt} satisfies $A_V$, while $P_{\mathit{fix}}$ requires accesses outside \texttt{ROOT} to be blocked. Since only the vulnerable execution performs the out-of-scope file access, \textsc{Oscar} identifies the out-of-scope file read as $r_V$.

\subsubsection{Realized Effect Reconstruction.}

After identifying where the effect begins, \textsc{Oscar} reconstructs the affected objects or states and additional security-relevant consequences. Following Finding~1, it represents the realized effect as
\[
E_V=(r_V,O_V,S_V),
\]
where $O_V$ denotes the objects or states affected by $r_V$, and $S_V$ denotes additional security-relevant values, states, or external outcomes. Starting from $r_V$, \textsc{Oscar} follows Resource and Data relations to identify runtime entities representing the affected objects or states $O_V$, and then follows Data, Resource, and Control relations to recover downstream security-relevant consequences $S_V$. If $r_V$ acting on $O_V$ already constitutes the complete effect, then $S_V=\varnothing$.

In the running example, the out-of-scope read acts on \texttt{/workspace/private/canary.txt} and obtains \texttt{VULN\_WITNESS\_7F3A}. Accordingly, $O_V=\{\texttt{/workspace/private/canary.txt}\}$ and $S_V=\{\texttt{VULN\_WITNESS\_7F3A}\}$.

\subsubsection{Effect Manifestation Localization.}
The final step determines how far the realized effect manifests in the Agent system.
Starting from the entities constituting $E_V$, \textsc{Oscar} follows their cross-layer provenance in $G_v$.
We regard an entity at a later layer as an effect representation when it remains
connected through runtime provenance to $r_V$, $O_V$, or $S_V$ and preserves the
corresponding security-relevant behavior, state, value, or outcome. 
This criterion does not require byte-level equality with the original runtime value: Host-side filtering, formatting, or projection may alter the concrete representation while preserving the same effect semantics. Conversely, merely carrying vulnerability-related diagnostic information is insufficient if the realized effect itself is no longer represented.
Following Finding~2, \textsc{Oscar} defines the maximum manifestation boundary as
\[
B(E_V)\in
\{
B_{\mathrm{Runtime}},
B_{\mathrm{Tool}},
B_{\mathrm{Observation}}
\}.
\]
Here, the Runtime boundary covers the Program and Resource layers, the Tool boundary covers the MCP and Host layers, and the Observation boundary corresponds to the model-visible Agent Observation. Accordingly, $B(E_V)=B_{\mathrm{Runtime}}$ if no effect representation enters the Tool Result, $B(E_V)=B_{\mathrm{Tool}}$ if it enters the Tool-return path but not the Agent Observation, and $B(E_V)=B_{\mathrm{Observation}}$ if it remains identifiable in the Observation. Host processing is treated as an intermediate stage rather than a separate maximum boundary.

In the running example, the value obtained by the out-of-scope file read enters the Tool Result, passes through MCP and Host processing, and remains identifiable in the Agent Observation. \textsc{Oscar} therefore assigns $B(E_V)=B_{\mathrm{Observation}}$.

\section{Evaluation}
\label{sec:experiments}

We evaluate \textsc{Oscar} from three aspects: its effectiveness in identifying realized vulnerability effects, its accuracy in localizing their maximum manifestation boundaries, and the contribution of its major design components.
Our experiments focus on the following Research Questions (RQs):

\noindent $\bullet$ \textbf{RQ1:} How accurately can \textsc{Oscar} determine whether a known OSS vulnerability forms a realized effect under a given Agent task and input?

\noindent $\bullet$ \textbf{RQ2:} How accurately can \textsc{Oscar} localize the maximum manifestation boundary of a realized vulnerability effect?

\noindent $\bullet$ \textbf{RQ3:} How do vulnerability-attribution evidence and graph representation contribute to \textsc{Oscar}'s analysis?

\subsection{Evaluation Setup}

\noindent\textbf{Implementation.}
We implemented \textsc{Oscar} in Python 3.12 following the three-stage workflow described in Section~\ref{sec:Oscar}.
For each case, \textsc{Oscar} extracts vulnerability trigger semantics $A_V$ and patch semantics $P_{\mathrm{fix}}$, executes the vulnerable and fixed versions, constructs their execution graphs, reconstructs
$E_V=(r_V,O_V,S_V)$, and determines $B(E_V)$.
The collected runtime trace covers program and resource events, Tool Results, Host-side representations, and model-visible Observations.

\noindent\textbf{Environment.}
All experiments were conducted on a server running Ubuntu 24.04.2 LTS, equipped with two 64-core AMD EPYC 7B13 processors, 188~GiB of memory, one 40~GiB NVIDIA A100 GPU, and one 80~GiB NVIDIA A100 GPU.
Docker 28.5.1 was used to isolate individual cases.
For paired executions, the Agent task, input, Host configuration, Tool set, container configuration, network settings, and external services were kept unchanged.
Only the target OSS version differed between the vulnerable and fixed executions.

\noindent\textbf{Baselines.}
We compare \textsc{Oscar} against three representative baselines: Code Reachability, AgentFuzz, and MCP-BiFlow.
\textit{Code Reachability} \cite{wu2023maven} checks whether the vulnerability-related code region is reachable from the current Tool entry.
\textit{AgentFuzz} \cite{liu2025agentfuzz} dynamically detects vulnerabilities in LLM agents by executing Agent tasks and observing vulnerability-triggering behavior.
\textit{MCP-BiFlow} \cite{hou2026mcpbiflow} detects unsafe data flows in MCP servers by recovering request-side and return-side propagation paths.
For case-level comparison, we normalize their outputs to \textsc{Effect Yes}, \textsc{Effect No}, or \textsc{Unassessed} according to each method's detection result.

\noindent\textbf{Dataset.}
We first collected popular MCP servers and Agent tools from public repositories and tool ecosystems, and used OSV to identify known vulnerabilities in these projects and their dependencies.
For each candidate, we manually examined the vulnerability advisory, affected versions, source code, and patch to confirm that the affected component lies on the Tool execution path.
We then constructed the corresponding Agent task and execution context using DeepSeek Harness \cite{deepseek-harness2026}, and manually annotated the effect label and, for positive cases, the realized effect and its maximum manifestation boundary based on the vulnerability semantics and execution context.

The preliminary study and evaluation cases are separated by project.
Development and test cases are further grouped by project-vulnerability family to avoid overlap between the two sets.
As shown in Table~\ref{tab:eval-dataset}, the final independent test set contains 50 cases, including 28 \textsc{Effect Yes} and 22 \textsc{Effect No}.
Among the 28 positive cases, 10 have a Runtime boundary, eight have a Tool boundary, and 10 have an Observation boundary.

\begin{table}[!ht]
    \centering
    \caption{Composition of the independent test set by effect label and maximum manifestation boundary.}
    \label{tab:eval-dataset}
    \resizebox{0.45\linewidth}{!}{
    \begin{tabular}{rrrrrrr}
        \toprule
        \multirow{2}{*}{Total}
        & \multicolumn{2}{c}{Effect Label}
        & \multicolumn{4}{c}{Boundary among \textsc{Effect Yes}} \\
        \cmidrule(lr){2-3}
        \cmidrule(lr){4-7}
        & Yes & No & Runtime & Tool & Observation \\
        \midrule
        50 & 28 & 22 & 10 & 8 & 10 \\
        \bottomrule
    \end{tabular}
    }
\end{table}

\noindent\textbf{Metrics.}
For realized effect identification, we report Precision, Recall, F1, Accuracy, and Coverage.
For boundary localization, we evaluate the 28 \textsc{Effect Yes} cases and report Boundary Accuracy, Boundary Coverage, Conditional Accuracy, and Macro-F1.
Boundary Accuracy is computed over all 28 positive cases, while Conditional Accuracy is computed only over cases for which a boundary is produced.
Boundary Macro-F1 is calculated over the three defined boundary classes: Runtime, Tool, and Observation.

\subsection{RQ1: Realized Effect Identification}
\label{sec:rq1}

To answer RQ1, we evaluate four methods on the 50-case independent test set, using \textsc{Effect Yes} as the positive class.
As shown in Table~\ref{tab:RQ1}, \textsc{Oscar} achieves 96.4\% Precision, Recall, and F1, with an Accuracy of 96.0\%.
Its F1 is 18.6, 7.9, and 22.9 percentage points higher than Code Reachability, AgentFuzz, and MCP-BiFlow, respectively.

\begin{table}[!ht]
    \centering
    \caption{Results of known OSS vulnerability effect identification.}
    \label{tab:RQ1}
    \resizebox{0.65\linewidth}{!}{
    \begin{tabular}{lrrrrrrrrr}
        \toprule
        Method & TP & FP & FN & TN & Prec. & Rec. & F1 & Acc. & Cov. \\
        \midrule
        Code Reachability
        & 28 & 16 & 0 & 6
        & 63.6 & 100.0 & 77.8 & 68.0 & 100.0 \\
        AgentFuzz
        & 23 & 1 & 5 & 21
        & 95.8 & 82.1 & 88.5 & 88.0 & 100.0 \\
        MCP-BiFlow
        & 18 & 3 & 10 & 19
        & 85.7 & 64.3 & 73.5 & 74.0 & 100.0 \\
        \textbf{\textsc{Oscar}}
        & \textbf{27} & \textbf{1} & \textbf{1} & \textbf{21}
        & \textbf{96.4} & \textbf{96.4} & \textbf{96.4}
        & \textbf{96.0} & \textbf{100.0} \\
        \bottomrule
    \end{tabular}
    }
\end{table}

\noindent\textbf{Code Reachability.}
Code Reachability identifies all 28 positive cases, but also produces 16 false positives.
Among them, nine cases reach the vulnerability-related code without satisfying the trigger condition.
The remaining seven execute vulnerability-related intermediate behavior without producing the expected change to the affected resource, value, or control state.
The absence of false negatives is expected because a realized effect necessarily executes the vulnerability-related path, while path reachability alone does not imply that the effect has occurred.

\noindent\textbf{AgentFuzz.}
AgentFuzz achieves 95.8\% Precision, 82.1\% Recall, and 88.5\% F1, with one false positive and five false negatives.
Compared with Code Reachability, AgentFuzz substantially reduces false positives by requiring runtime evidence of vulnerability triggering.
Its only false positive occurs in a case where a vulnerability-related control signal is observed, but no corresponding change to the target resource, sensitive value, or persistent state follows.
Thus, the trigger confirms activation of vulnerability-related behavior without establishing that a concrete effect has been realized.

The five false negatives exhibit the opposite pattern.
In these cases, the vulnerability changes an external resource, file state, or persistent state without consistently producing a return value, exception, crash, or other trigger signal captured by AgentFuzz.
The effect is therefore realized in the execution environment but is not reflected in the runtime signals used for triggering detection.
These results show that AgentFuzz effectively filters many reachability-only cases, but may miss effects whose primary manifestation is an environmental state change.

\noindent\textbf{MCP-BiFlow.}
MCP-BiFlow detects all 10 Observation-boundary and all eight Tool-boundary effects, but misses all 10 Runtime-boundary cases.
In these cases, the vulnerability has already changed a file, process, network interaction, or persistent state, but no representation of the effect enters the Tool Result.
Such effects therefore do not form a return-side carrier that can be followed through subsequent data propagation.
The three false positives expose a different limitation.
Although a source-to-sink path exists, the propagated value does not by itself establish that the runtime consequence of the target CVE has been realized.
Thus, propagation evidence depends both on the presence of a propagating carrier and on its attribution to the vulnerability effect.

\noindent\textbf{\textsc{Oscar}.}
\textsc{Oscar} correctly identifies 27 of the 28 positive cases and rejects 21 of the 22 negative cases.
It identifies all 10 Observation-boundary cases, all eight Tool-boundary cases, and nine of the 10 Runtime-boundary cases.
Among the 22 negative cases, \textsc{Oscar} correctly rejects all nine cases in which vulnerable code is reached without a realized effect, all six cases in which the vulnerability is not triggered or the relevant path is unreachable, and six of the seven cases in which a vulnerability-related mechanism executes without producing a concrete consequence.

The only false positive occurs when an intermediate control behavior is incorrectly attributed to the target vulnerability effect.
Although vulnerability-related control transitions and diagnostic signals are observed, no corresponding resource change, sensitive value, persistent-state modification, or other concrete consequence is produced.
The only false negative results from an incomplete external-state witness.
The effect occurs in an external resource or persistent state, but the state change cannot be reliably linked to the corresponding Tool invocation and execution instance, preventing \textsc{Oscar} from confirming its attribution to the target CVE.

Overall, the error patterns reflect the different evidence captured by the three baselines.
Code Reachability establishes execution of a vulnerability-related path, AgentFuzz detects vulnerability-triggering behavior, and MCP-BiFlow follows values that propagate across Tool boundaries.
For known OSS vulnerabilities, each provides useful evidence, but none alone establishes both that the expected runtime consequence has occurred and that it is attributable to the target CVE.
\textsc{Oscar} combines vulnerability-specific semantics, paired execution, and cross-layer runtime evidence to distinguish path reachability and mechanism activation from a realized vulnerability effect.

\begin{titleEnv}
\textit{\textbf{Answer to RQ1}: \textsc{Oscar} achieves 96.4\% F1 and 96.0\% Accuracy in identifying realized OSS vulnerability effects.
Its advantage comes from confirming both the concrete runtime consequence and its attribution to the target CVE, rather than treating reachability, triggering, or propagation alone as sufficient evidence of effect.}
\end{titleEnv}

\subsection{RQ2: Maximum Manifestation Boundary Localization}
\label{sec:rq2}

To answer RQ2, we evaluate \textsc{Oscar} on the 28 cases in which a realized effect is present.
The set contains 10 Runtime-, 8 Tool-, and 10 Observation-boundary effects.
\textsc{Oscar} produces a boundary for 27 cases and correctly localizes 24, yielding 85.7\% Boundary Accuracy, 96.4\% Boundary Coverage, 88.9\% Conditional Accuracy, and 87.1\% Macro-F1.

\begin{table}[!ht]
    \centering
    \caption{Maximum manifestation boundary localization results.}
    \label{tab:RQ2}
    \resizebox{0.75\linewidth}{!}{
    \begin{tabular}{lrrrrrrrr}
        \toprule
        \multirow{2}{*}{Ground Truth}
        & \multicolumn{4}{c}{Predicted Boundary}
        & \multirow{2}{*}{Cases}
        & \multirow{2}{*}{Prec.}
        & \multirow{2}{*}{Rec.}
        & \multirow{2}{*}{F1} \\
        \cmidrule(lr){2-5}
        & Runtime & Tool & Observation & Unassessed
        & & & & \\
        \midrule
        Runtime
        & 8 & 1 & 0 & 1
        & 10 & 100.0 & 80.0 & 88.9 \\

        Tool
        & 0 & 7 & 1 & 0
        & 8 & 77.8 & 87.5 & 82.4 \\

        Observation
        & 0 & 1 & 9 & 0
        & 10 & 90.0 & 90.0 & 90.0 \\
        \midrule
        Macro
        & -- & -- & -- & --
        & 28 & 89.3 & 85.8 & 87.1 \\
        \bottomrule
    \end{tabular}
    }
\end{table}

\noindent\textbf{Runtime boundary.}
\textsc{Oscar} correctly localizes 8 of the 10 Runtime cases.
In these cases, the realized effect is confirmed from program operations or resource-state changes, while no representation preserving the effect appears in the Tool Result.
One case is incorrectly localized to Tool because vulnerability-related diagnostic information in the Tool Result is mistaken for an effect representation.
The other case is marked \textsc{Unassessed} because the external-state evidence is incomplete.
This error shows that vulnerability-related information in a Tool Result does not necessarily represent the realized effect itself.

\noindent\textbf{Tool boundary.}
\textsc{Oscar} correctly localizes 7 of the 8 Tool cases, with the lowest F1 among the three boundary classes.
A Tool-boundary effect must remain identifiable in the Tool Result but no longer be preserved after Host processing.
The only error occurs when Host-side processing is incorrectly interpreted as preserving the effect into the model-visible Observation, causing the boundary to be localized to Observation.
This case highlights the difficulty of determining whether an effect representation survives Host-side transformation rather than merely appearing in related intermediate data.

\noindent\textbf{Observation boundary.}
\textsc{Oscar} correctly localizes 9 of the 10 Observation cases.
These cases require the analysis to track the effect from the Tool Result through Host-side transformations to the content actually included in the model request.
The only error occurs when a nested Host transformation changes the representation structure and breaks its correspondence with the original effect, causing the analysis to stop at Tool.
Although the effect remains represented in the model-visible content, \textsc{Oscar} fails to recover this cross-layer correspondence.

The three mislocalizations all arise when determining whether vulnerability-related information at a later stage still represents the same realized effect.
No Runtime case is directly localized to Observation, and no Observation case is reduced to Runtime.
The remaining errors are therefore concentrated at cross-layer representation changes.

These results show that boundary localization cannot rely on the presence of related values alone.
A Tool Result may contain vulnerability-related diagnostic information without carrying the realized effect, while the same effect may survive Host processing even when its representation changes.
Accurate localization therefore requires tracking the security-relevant semantics of the effect across execution layers.

\begin{titleEnv}
\textit{\textbf{Answer to RQ2}: \textsc{Oscar} correctly localizes 24 of 28 realized effects, achieving 85.7\% Boundary Accuracy.
The results show that \textsc{Oscar} can recover how far a realized OSS vulnerability effect propagates through the Agent execution stack, including cases where its representation changes across Tool and Host processing.
}
\end{titleEnv}

\subsection{RQ3: Ablation Study}
\label{sec:rq3}

We conduct two ablation studies.
The first removes $P_{\mathrm{fix}}$ or the fixed-version execution to evaluate vulnerability attribution.
The second replaces \textsc{Oscar}'s cross-layer execution graph with Raw Trace and Generic Provenance Graph to evaluate the role of evidence representation in realized effect identification and boundary localization.

\subsubsection{Vulnerability-Attribution Evidence.}
\label{sec:rq3-attribution}

Table~\ref{tab:attribution-ablation} compares the full \textsc{Oscar} pipeline with two variants.
\textit{w/o $P_{\mathrm{fix}}$} removes patch semantics while retaining the vulnerable and fixed executions.
\textit{w/o Fixed Run} retains $A_V$, $P_{\mathrm{fix}}$, and the vulnerable execution, but removes the fixed-version execution.

\begin{table}[!ht]
    \centering
    \caption{Ablation of vulnerability-attribution evidence.}
    \label{tab:attribution-ablation}
    \resizebox{0.6\linewidth}{!}{
    \begin{tabular}{lrrrrrrrr}
        \toprule
        Variant & TP & FP & FN & TN & Prec. & Rec. & F1 & Acc. \\
        \midrule
        w/o $P_{\mathrm{fix}}$
        & 25 & 6 & 3 & 16
        & 80.6 & 89.3 & 84.7 & 82.0 \\
        w/o Fixed Run
        & 26 & 5 & 2 & 17
        & 83.9 & 92.9 & 88.1 & 86.0 \\
                \textbf{Full \textsc{Oscar}}
        & \textbf{27} & \textbf{1} & \textbf{1} & \textbf{21}
        & \textbf{96.4} & \textbf{96.4} & \textbf{96.4} & \textbf{96.0} \\
        \bottomrule
    \end{tabular}
    }
\end{table}

\noindent\textbf{Without Patch Semantics.}
Removing $P_{\mathrm{fix}}$ reduces F1 from 96.4\% to 84.7\%.
The variant produces five additional false positives and two additional false negatives.
Without patch semantics, version-specific differences unrelated to the target vulnerability and behaviors resembling the vulnerable operation are more easily attributed to the target CVE.
At the same time, some true effects can no longer be linked clearly to the behavior corrected by the patch.

\noindent\textbf{Without Fixed-Version Execution.}
Removing the fixed-version execution reduces F1 from 96.4\% to 88.1\%, with four additional false positives and one additional false negative.
Although $P_{\mathrm{fix}}$ still describes the behavior corrected by the patch, the analysis can no longer verify whether the corresponding runtime behavior disappears under the fixed implementation.
As a result, behaviors that are consistent with the vulnerability semantics but are not specific to the vulnerable version are more likely to be attributed to the target CVE.

This result shows that $P_{\mathrm{fix}}$ provides the semantic criterion for attribution, while the fixed execution provides the runtime evidence needed to validate that criterion.

\subsubsection{Evidence Representation.}
\label{sec:rq3-representation}

We next compare three representations of the same underlying runtime evidence: \textit{Raw Trace}, \textit{Generic Provenance Graph}, and \textit{Full \textsc{Oscar}}.
All three variants use the same vulnerable and fixed executions, source code, patch information, candidate runtime evidence, Host evidence, repeated runs, and decision thresholds.
Only the representation used to organize and relate this evidence differs.

\textit{Raw Trace} preserves the ordered runtime records without explicit dependency relations.
\textit{Generic Provenance Graph} models each Agent run as a graph over execution steps, with execution edges and dependency edges that capture step ordering and
inter-step dependencies.
\textit{Full \textsc{Oscar}} uses the cross-layer execution graph defined in Section~\ref{sec:Oscar}, further representing the program, resource, Tool, Host, and Agent entities and relations required for effect reconstruction and manifestation analysis.

\begin{figure*}[t]
    \centering
    \includegraphics[width=\textwidth]{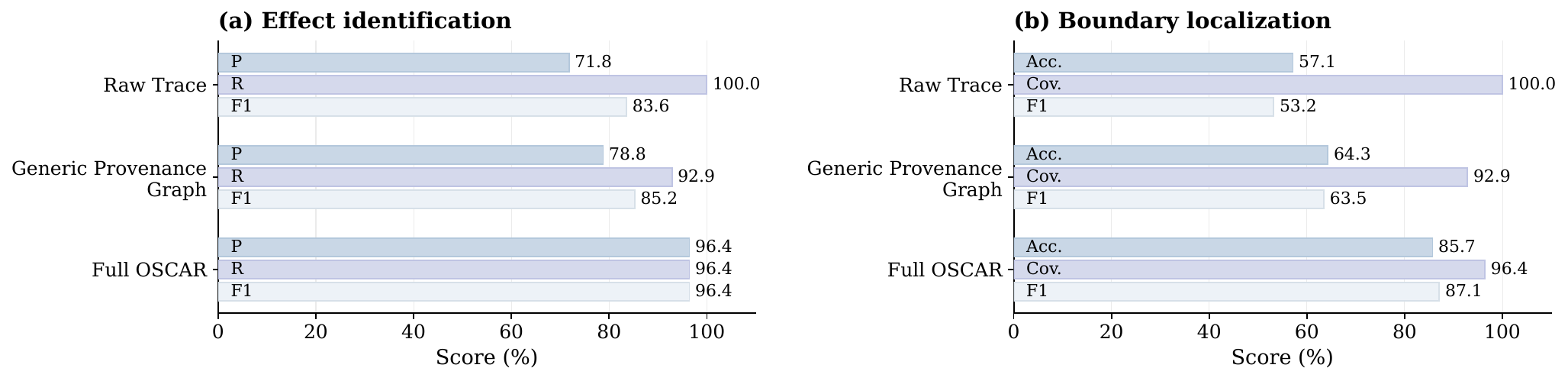}
    \caption{Effect of evidence representation on realized effect identification and boundary localization.}
    \label{fig:rq3}
\end{figure*}

\noindent\textbf{Raw Trace.}
As shown in Figure~\ref{fig:rq3}, Raw Trace detects all 28 positive cases, but also classifies 11 of the 22 negative cases as \textsc{Effect Yes}, resulting in an F1 of 83.6\%.
The trace contains the relevant runtime events together with ordinary return values, error messages, protocol fields, and other execution differences.
Without explicit provenance structure, it is difficult to determine which of these observations belong to the realized effect of the target vulnerability.

The same limitation is more pronounced in boundary localization.
Raw Trace produces a boundary for all 28 positive cases, but localizes only 16 correctly.
In particular, 6 Runtime cases and 6 Tool cases are incorrectly promoted to Observation because related differences also appear at later execution stages.
A related value or difference appearing later in the trace is therefore insufficient to establish that the realized effect itself has propagated to that stage.

\noindent\textbf{Generic Provenance Graph.}
Generic Provenance Graph reduces false positives from 11 to 7 and increases Effect F1 from 83.6\% to 85.2\%.
Call identity, value correspondence, and resource provenance help reject some runtime differences that lack a consistent provenance relation.
However, two positive cases are missed, and the overall improvement remains limited.
Generic provenance establishes relations among runtime events, but does not explicitly encode their roles in $E_V=(r_V,O_V,S_V)$ or whether a later representation preserves the security-relevant semantics of the realized effect.

A similar limitation appears in boundary localization.
Generic Provenance Graph produces boundaries for 26 of the 28 positive cases and correctly localizes 18.
It correctly localizes 8 of the 10 Runtime cases, but only 2 of the 8 Tool cases.
The remaining 6 Tool cases are promoted to Observation.
Two Observation cases remain \textsc{Unassessed} because the provenance evidence is insufficient to connect their later representations to the realized effect.

\noindent\textbf{Full \textsc{Oscar}.}
Full \textsc{Oscar} achieves 96.4\% Effect F1 and 85.7\% Boundary Accuracy.
Unlike generic provenance, the \textsc{Oscar} graph represents runtime entities according to their roles in the reconstructed effect and explicitly connects program operations, affected resources, Tool Results, Host processing, and Agent Observations.
This representation enables the analysis to distinguish information that merely appears at a later stage from a representation that still preserves the security-relevant semantics of the same realized effect.

The two ablation experiments isolate two different requirements of \textsc{Oscar}.
Removing $P_{\mathrm{fix}}$ or the fixed-version execution weakens the attribution of observed runtime behavior to the target vulnerability.
Replacing the \textsc{Oscar} graph with less structured runtime representations weakens the reconstruction and cross-layer tracking of the realized effect.
Together, the results show that accurate OSS vulnerability analysis requires both vulnerability-specific attribution and effect-aware runtime representation.

\begin{titleEnv}
\textit{\textbf{Answer to RQ3}: \textsc{Oscar} benefits from two distinct components.
Vulnerability-specific attribution, provided by patch semantics and fixed-version execution, distinguishes effects of the target vulnerability from other runtime behavior.
The cross-layer execution graph then reconstructs and tracks those effects across the Agent execution.}
\end{titleEnv}

\section{Real-World Applicability}
\label{sec:realworld}

We evaluate \textsc{Oscar} on ARIS (Auto-Research-In-Sleep)~\cite{yang2026aris}, an open-source autonomous research harness that coordinates multiple agents for tasks such as literature review, experiment execution, and paper writing.

\subsection{Real-World Evaluation Setup}

ARIS is well suited for this evaluation because it supports multiple Agent Hosts within a unified workflow.
As of October 2, 2026, its GitHub repository has 16,901 stars and 1,428 forks~\cite{aris_repo}.
To ensure reproducibility, we pin ARIS to commit \texttt{2132036} \cite{aris_commit} and deploy five Agent Hosts: DeepSeek Harness, Qwen-Agent~\cite{qwenagent}, PraisonAI~\cite{praisonai}, Goose~\cite{goose}, and OpenCode~\cite{opencode}.
These Hosts differ in Tool invocation, MCP integration, session representation, and model-request construction, allowing us to evaluate \textsc{Oscar} across heterogeneous Agent runtime implementations.

We first use OSV-Scanner~2.6.0~\cite{osvscanner} to scan the resolved dependency environments of the five Hosts.
In total, we analyze 5,941 dependency entries and identify 117 vulnerable package-version entries, corresponding to 303 unique CVE identifiers.
We then analyze code reachability and retain 25 reachable vulnerability instances whose vulnerable functionality can be exercised through normal Agent functionality.

For each instance, we construct an Agent task following the normal usage described in the corresponding Host README and execute it using the Host's native Tool interface, runtime environment, and configuration. Under the same Agent task, Tool input, Host configuration, and external environment, we run both the vulnerable and fixed versions of the target OSS component, repeating each version three times. \textsc{Oscar} then identifies realized vulnerability effects and localizes their maximum manifestation boundaries. In parallel, we record complete runtime traces for manual validation, including vulnerability-relevant program operations, resource/state changes, Tool Results, host-processed representations, and actual model requests.

\subsection{Results}

Among the 25 reachable vulnerability instances, \textsc{Oscar} identifies realized vulnerability effects in nine cases and successfully localizes their maximum manifestation boundaries, as shown in Table~\ref{tab:realworld-cases}.
For the remaining cases, \textsc{Oscar} either finds no realized effect attributable to the target vulnerability or lacks sufficient evidence to complete the attribution.

\begin{table}[!ht]
    \centering
    \caption{Real-world vulnerability cases identified by \textsc{Oscar}.}
    \label{tab:realworld-cases}
    \resizebox{0.95\linewidth}{!}{
    \begin{tabular}{llll}
    \toprule
    \textbf{Agent Host} & \textbf{CVE / Component} & \textbf{Observed Realized Effect} & \textbf{Boundary} \\
    \midrule
    DeepSeek Harness &
    CVE-2026-33989 / mobile-mcp &
    Screenshot file created outside intended directory &
    Runtime \\

    Qwen-Agent &
    CVE-2024-47081 / Requests &
    \texttt{.netrc}-derived credential enters model request &
    Observation \\

    Qwen-Agent &
    CVE-2026-33989 / mobile-mcp &
    Screenshot file created outside intended directory &
    Runtime \\

    PraisonAI &
    CVE-2024-47081 / Requests &
    \texttt{.netrc}-derived credential enters model request &
    Observation \\

    PraisonAI &
    CVE-2026-33989 / mobile-mcp &
    Screenshot file created outside intended directory &
    Runtime \\

    Goose &
    CVE-2026-32871 / FastMCP &
    Unauthorized administrator response enters model request &
    Observation \\

    Goose &
    CVE-2025-11849 / Mammoth &
    Document conversion causes out-of-scope file-state modification &
    Runtime \\

    OpenCode &
    CVE-2026-44243 / GitPython &
    Illegal ref path is written and later returned by a diff Tool &
    Observation \\

    OpenCode &
    CVE-2024-47081 / Requests &
    \texttt{.netrc}-derived Authorization enters model request &
    Observation \\
    \bottomrule
    \end{tabular}
    }
\end{table}

We manually inspect the complete runtime traces of the nine positive cases.
The inspection confirms that all nine cases indeed produce realized vulnerability effects corresponding to the target CVEs, and that \textsc{Oscar}'s recovered effects and manifestation boundaries are consistent with the runtime evidence.
Four cases have Runtime as their maximum manifestation boundary, while five reach Observation.

The nine cases exhibit two manifestation patterns.
In the first, the vulnerability effect changes a concrete resource or state in the underlying runtime, but no representation of the effect enters the Tool Result.
In the second, the effect propagates through the Tool Result and Host processing and eventually reaches the model-visible Observation.
These results demonstrate that \textsc{Oscar} can identify realized OSS vulnerability effects across heterogeneous Agent runtime implementations and accurately localize their cross-layer manifestation boundaries.

\subsection{Representative Cases}

We select two cases with different manifestation boundaries to illustrate \textsc{Oscar}'s analysis in real-world Agent executions.

\noindent\textbf{Case 1: Runtime-Bound File-System Effect.}
On Qwen-Agent, the Agent executes a normal screenshot task:

\begin{tcolorbox}[
  colback=gray!8,
  colframe=gray!35,
  boxrule=0.5pt,
  arc=2pt,
  left=6pt,
  right=6pt,
  top=5pt,
  bottom=5pt
]
\texttt{Capture a screenshot of the test device and save it to the output location specified in the job configuration.}
\end{tcolorbox}

The output path is set to \texttt{../escape.png}, satisfying the trigger semantics of CVE-2026-33989 in mobile-mcp.
The vulnerable mobile-mcp~0.0.48 accepts this path and creates a screenshot file outside the intended directory, whereas the fixed version~0.0.49 rejects the same input before the write occurs.
\textsc{Oscar} identifies the out-of-scope file write as the realized vulnerability effect, which is independently confirmed through file-system inspection.

The Tool Result contains no representation of this file-system effect.
\textsc{Oscar} therefore localizes the maximum manifestation boundary at Runtime.
This case demonstrates that a vulnerability can produce a concrete resource-state change without manifesting in the Tool Result.

\noindent\textbf{Case 2: Observation-Bound Credential Effect.}
On OpenCode, the Agent executes a normal remote-service query task:

\begin{tcolorbox}[
  colback=gray!8,
  colframe=gray!35,
  boxrule=0.5pt,
  arc=2pt,
  left=6pt,
  right=6pt,
  top=5pt,
  bottom=5pt
]
\texttt{Retrieve and summarize the latest ServiceNow incident.}
\end{tcolorbox}

The experimental environment contains a \texttt{.netrc} credential, the target ServiceNow authority, and a controlled HTTP endpoint, satisfying the trigger conditions of CVE-2024-47081 in Requests.
In vulnerable Requests~2.32.3, the program performs an unintended \texttt{.netrc} credential selection and sends the resulting Authorization value to the controlled HTTP service.
Under the same execution conditions, fixed Requests~2.32.4 no longer exhibits this behavior.

The vulnerability-derived value subsequently enters the Tool Result.
OpenCode stores the Tool Result as a session event and incorporates it into the next model request.
Inspection of the actual model request confirms that the same Authorization value remains identifiable.
\textsc{Oscar} therefore localizes the maximum manifestation boundary at Observation.
This case demonstrates an Observation-bound effect whose representation persists through the Tool Result and Host processing into the model-visible Observation.

\begin{titleEnv}
\textit{\textbf{Summary}: Among the 25 reachable vulnerability instances, \textsc{Oscar} identifies nine realized vulnerability effects and localizes their manifestation boundaries. Manual inspection confirms all nine results: four terminate at Runtime and five reach Observation.}
\end{titleEnv}

\section{Discussion}
\label{sec:discussion}

\textbf{Scope of Analysis.}
\textsc{Oscar} analyzes the realized effect of a known OSS vulnerability under a given Agent task, input, and execution context. 
A negative result therefore means that no attributable security-relevant effect is established for the analyzed execution.
It does not imply that the vulnerability is generally harmless or unreachable under other inputs.

\noindent\textbf{Vulnerability Attribution.}
Attributing runtime changes to a target CVE is difficult because vulnerable and fixed versions may contain unrelated differences.
\textsc{Oscar} therefore combines vulnerability trigger semantics $A_V$, patch semantics $P_{fix}$, and paired vulnerable/fixed executions. Ablation results support this design: removing $P_{fix}$ or the fixed execution reduces realized effect identification F1 from 96.4\% to 84.7\% and 88.1\%, respectively.
Attribution may still be affected by incomplete vulnerability descriptions or unrelated changes in fixing commits.

\noindent\textbf{Runtime Evidence Completeness.}
Vulnerability effects may appear in program state, external resources, or persistent environments without explicit return values. \textsc{Oscar} captures program operations, resource interactions, runtime values, and external-state changes in a cross-layer execution graph, enabling detection of Runtime-boundary effects that return-side analysis may miss. Effect confirmation nevertheless requires an observable witness linked to the relevant execution. In our evaluation, the only false negative arises from an external-state change whose provenance cannot be reliably associated with the corresponding Tool invocation.

\noindent\textbf{Cross-Layer Effect Correspondence.}
A realized effect may change representation as it propagates through Tool Results, Host processing, and Agent Observations. \textsc{Oscar} preserves runtime provenance from the root change to downstream representations using an effect-aware cross-layer execution graph, rather than relying only on value similarity or co-occurrence. This improves boundary localization over Raw Trace and Generic Provenance representations. Complex Host-side transformations can still obscure whether downstream representations preserve the same security-relevant effect, accounting for the remaining RQ2 localization errors.

\noindent\textbf{Generalizability and Realism.}
\textsc{Oscar}'s cross-layer execution graph is extensible to evolving Agent runtimes by adding or refining node and relation types without changing the effect reconstruction and localization procedure. Evaluation across five heterogeneous Agent Hosts provides initial evidence for this portability.
Since naturally occurring Agent workloads and user inputs are unavailable, we construct vulnerability-relevant tasks to evaluate \textsc{Oscar}'s analysis capability. Our real-world study further follows documented usage and native execution environments while exercising vulnerability-relevant conditions. The results therefore support portability across Agent runtimes, rather than vulnerability triggering frequency in natural workloads.

\section{Related Work}
\label{sec:related-work}

\subsection{Vulnerability Analysis of Open Source Software Components}

\noindent\textbf{OSS Component and Vulnerable Code Analysis.}
OSSPolice identifies reused components and versions~\cite{duan2017osspolice}, CENTRIS detects modified and nested reuse~\cite{woo2021centris}, and OSSFP uses function fingerprints~\cite{wu2023ossfp}. VulSCA adds code-level vulnerability confirmation~\cite{hu2026vulsca}, while code-centric analysis examines vulnerable code in downstream applications~\cite{ponta2018beyond}. Graph-based methods model program dependencies with heterogeneous code graphs~\cite{zhang2024dshgt} or use code-property-graph-guided slicing~\cite{lekssays2025llmxcpg}. These studies cover component identification, vulnerable-code detection, and related analysis~\cite{lin2025ossurvey}, but do not capture the concrete runtime effects of vulnerabilities.

\noindent\textbf{Vulnerability Trigger and Patch Analysis.}
PatchFinder traces vulnerabilities to fixing commits~\cite{li2024patchfinder}, E-SPI identifies security patches from commit structure~\cite{wu2022espi}, and GraphSPD represents patch changes with code property graphs~\cite{wang2023graphspd}. PS$^3$ derives semantic signatures for patch-presence testing~\cite{zhan2024ps3}, while Robin distinguishes vulnerable and patched behavior~\cite{yang2023robin}. P$^3$ jointly analyzes pre- and post-patch versions~\cite{sharma2025p3}. These studies recover vulnerability triggers and fix actions~\cite{gao2025whyhow}, but do not reconstruct the resulting runtime effect or trace it beyond the vulnerable program.

\subsection{Vulnerability Analysis in Agent and MCP Systems}

\noindent\textbf{Program Analysis for LLM-Integrated Systems.}
LLMSmith combines call-chain analysis with execution validation~\cite{liu2023llmsmith}, and TaintP2X tracks model-generated data to sensitive operations~\cite{he2026taintp2x}. AgentFlow models Agent components and control/data dependencies~\cite{wang2026agentflow}. Agent-BOM combines system structure with runtime semantic states~\cite{li2026agentbom}, while Agentproof checks workflow-level safety properties~\cite{xavier2026agentproof}. These methods extend program analysis to model--program interactions, but do not trace the effect of a specific OSS vulnerability across Agent-system layers.

\noindent\textbf{Runtime Testing of Agent Systems.}
ToolEmu evaluates Agent risks in an emulated environment~\cite{ruan2024toolemu}. AgentFuzz uses directed fuzzing~\cite{liu2025agentfuzz}, AGENTVIGIL tests indirect prompt injection~\cite{wang2025agentvigil}, MELON compares original and masked re-executions~\cite{zhu2025melon}, and ChainFuzzer fuzzes multi-tool execution chains~\cite{wu2026chainfuzzer}. These approaches validate concrete Agent security behaviors, but do not attribute them to known OSS vulnerabilities or reconstruct their effects.

\noindent\textbf{Runtime Provenance and Effect Tracing.}
KAIROS builds whole-system provenance graphs~\cite{cheng2024kairos}, R-CAID performs provenance-based root-cause analysis~\cite{goyal2024rcaid}, and ORTHRUS reconstructs dependency paths~\cite{jiang2025orthrus}. MAGIC~\cite{jia2024magic} and FLASH~\cite{rehman2024flash} learn provenance-graph representations. SHADEWATCHER models audit interactions~\cite{zeng2022shadewatcher}, while DEPIMPACT propagates dependency impact~\cite{fang2022depimpact}. Provenance-based auditing reasons over execution dependencies~\cite{inam2023sokprovenance}, but mainly focuses on system-level events rather than vulnerability-specific effects~\cite{hossain2017sleuth}.

\noindent\textbf{Security Analysis of MCP Systems.}
MCP Safety Audit evaluates MCP-connected workflows~\cite{radosevich2025mcpsafetyaudit}, while large-scale server analysis studies MCP security~\cite{hasan2026mcpfirstglance}. MCP-BiFlow performs bidirectional data-flow analysis~\cite{hou2026mcpbiflow}. DCIChecker checks description--implementation inconsistencies~\cite{shi2026dcichecker}, MCPZoo supports runtime validation~\cite{chen2026mcpzoo}, and MindGuard analyzes tool poisoning with decision-dependence graphs~\cite{wang2025mindguard}. These methods focus on the Agent-Tool boundary, but do not establish the end-to-end connection between known OSS vulnerabilities and their runtime effects in Agent systems.

\section{Conclusion}
\label{sec:conclusion}
This paper studies how known OSS vulnerabilities manifest in agent executions. We present \textsc{Oscar}, a runtime analysis framework that identifies the security-relevant effect actually caused by a target vulnerability and determines its maximum manifestation boundary. \textsc{Oscar} combines vulnerability-specific semantics with vulnerable/fixed paired execution to attribute runtime changes to the target vulnerability, and uses cross-layer runtime provenance to recover their manifestations across the Agent stack. Our evaluation shows that \textsc{Oscar} achieves 96.4\% F1 and 96.0\% accuracy in realized vulnerability effect identification and correctly localizes 24 of 28 maximum manifestation boundaries. Evaluation on five Agent Hosts in the real-world ARIS framework further demonstrates that \textsc{Oscar} can analyze vulnerability effects that either remain within the underlying runtime or reach model-visible Observations across heterogeneous agent runtimes.

\section*{Data Availability}

An anonymized replication package with the materials needed to reproduce our results is available at \url{https://anonymous.4open.science/r/OSCAR-33E5/}.

\bibliographystyle{ACM-Reference-Format}
\bibliography{references}

\end{document}